\documentclass[twocolumn,showpacs,preprintnumbers,amsmath,amssymb]{revtex4}
\usepackage{graphicx}
\usepackage{dcolumn}
\usepackage{bm}
\def\bea {\begin{eqnarray}}
\def\eea {\end{eqnarray}}

\def\be {\begin{equation}}
\def\ee {\end{equation}}

\begin{document}

\title{Dragging  Heavy Quarks in Quark Gluon Plasma at the 
Large Hadron Collider}
\author{Santosh K Das, Jan-e Alam and Payal Mohanty}
\medskip
\affiliation{Variable Energy Cyclotron Centre, 1/AF, Bidhan Nagar , 
Kolkata - 700064}
\date{\today}
\begin{abstract}
The drag and diffusion coefficients of charm and bottom quarks propagating
through quark gluon plasma (QGP) have been evaluated for conditions
relevant to  nuclear collisions at Large Hadron Collider (LHC).  
The  dead cone and Landau-Pomeronchuk-Migdal (LPM) effects on 
radiative energy loss of heavy quarks have been considered. 
Both radiative and collisional processes of energy loss are included
in the {\it effective} drag and diffusion coefficients. 
With these effective transport coefficients we solve the Fokker Plank (FP)
equation for the heavy quarks executing Brownian motion
in the QGP. The solution of the FP equation has been used to
evaluate the nuclear suppression factor, $R_{\mathrm AA}$ for the
non-photonic single electron spectra resulting from the semi-leptonic decays 
of hadrons containing charm and bottom quarks. 
The effects of mass on $R_{\mathrm AA}$ has also been highlighted.
\end{abstract}

\maketitle

\section{Introduction}
Energy dissipation of heavy quarks in QCD matter is considered as 
one of the most  promising probe
for the quark gluon plasma (QGP) diagnostics. 
The energy loss of energetic 
heavy quarks ($Q$) while propagating through the QGP medium
is manifested in the suppression of 
heavy flavoured hadrons at high transverse momentum ($p_T$). 
The depletion of high $p_T$ hadrons ($D$ and $B$ mesons) 
produced in Nucleus + Nucleus collisions with respect to those
produced in proton + proton (pp) collisions has been measured 
experimentally~\cite{stare,phenixelat,phenixe} 
through their semi-leptonic decays.
The two main processes which cause this depletion are  
(i) elastic collisions 
and (ii) the bremsstrahlung or radiative loss due to the
interaction of the  heavy quarks with the quarks, 
anti-quarks and gluons in the thermal bath created in heavy ion collisions.

The importance of collisional energy loss in QGP diagnostics was
discussed first by Bjorken~\cite{bjorken}. 
The calculations of elastic loss  were performed with improved 
techniques~\cite{TG,peshier}
and its importance were highlighted subsequently~\cite{akdm,MT} in
heavy ion collisions. 
The collisional energy loss of heavy quarks~\cite{braaten}
has gained importance recently in view of the measured
nuclear suppression in the $p_T$ spectra of 
non-photonic single electrons. 
Several ingredients like inclusions of non-perturbative
contributions from the quasi-hadronic bound state~\cite{hvh},
3-body scattering effects~\cite{ko},
the dissociation of heavy mesons due to its 
interaction with the partons in the
thermal medium~\cite{adil}  and employment of running coupling
constants and realistic Debye mass~\cite{gossiaux}
have been proposed to improve the description of the  experimental data.
Wicks {\it et al.}~\cite{wicks} showed that the inclusion of both elastic 
and inelastic collisions and the path length fluctuation reduces the 
gap between the theoretical and experimental results.

The energy loss of energetic partons by radiation is a field of 
high current interest ~\cite{glv,zoww,bdps,salgado,revieweloss}.
For mass  dependence of energy loss due to radiative processes 
Dokshitzer and Kharzeev~\cite{DK} argue that
heavy quarks will lose much less energy than light quarks due to 
dead cone effects~\cite{rkellis}.  However, 
Aurenche and Zakharov claim that the radiative process has an anomalous mass
dependence~\cite{zakharov} due to the finite size of the QGP 
which leads to  small difference in energy loss between
a heavy and a light quarks. The mass dependence of the 
transverse momentum spectrum of the radiated gluons from 
the heavy quarks is studied in~\cite{asw}.  They found that
the medium induced gluon radiation fills up the dead cone with a reduced
magnitude at large gluon energies compared to the radiation 
from a light quarks. For  high  energy heavy quarks
the effects of the dead cone, however,  reduces because the magnitude 
of the angle 
forbidden for gluon emission behave as 
$\sim$ heavy quark mass/energy~\cite{wang}. From the study of
the mass dependence of the radiative loss it is shown 
in~\cite{adsw} that the very energetic charm (not the bottom)
quarks behave like massless
partons.  Although the authors in~\cite{roy} concluded that 
the suppression of radiative loss for heavy quarks is 
due to dead cone effects but it will be fair to state
that the issue is not settled yet.
 
The other mechanism which can affect the radiative loss is the 
LPM effect~\cite{wg} which 
depends on the relative magnitude of two time
scales of the system~\cite{klein}: the formation time  
($\tau_F$) and the mean scattering time scale ($\tau_c$) 
of the emitted gluons.
If $\tau_F\,>\,\tau_c$ then LPM suppression
will be effective.  The LPM effect is built-in in the expression for
radiative energy loss of heavy quarks derived in~\cite{asw,wang,adsw,dg}.
In contrast to those, in the present work  we will separately introduce 
the LPM effects in the energy loss formula.

The successes of the relativistic hydrodynamical 
model (see~\cite{pasi,teaney} for review) 
in describing the host of experimental results from 
Relativistic Heavy Ion Collider (RHIC)~\cite{npa2005}
indicate that the thermalization might have taken place in the system
of quarks and gluons formed after the nuclear collisions. 
The strong final state interaction of high energy partons  with
the QGP {\it i.e.} the observed jet quenching~\cite{phenix1,star1} and
the large elliptic flow ($v_2$)~\cite{phenix2,star2} in Au+Au collisions at RHIC
indicate the possibility of fast equilibration.
On the one hand 
the experimental data indicate early thermalization time $\sim 0.6$ fm/c
~\cite{arnold1} on the other hand the pQCD based calculations give a 
thermalization time $\sim 2.5$ fm/c~\cite{baier}(see also~{\cite{raju}). 
The gap between
these two time scales suggests that  the non-perturbative effects 
play a crucial role in achieving thermalization. It has also been
pointed out that the instabilities~\cite{mrowczynski,romatschke,arnold2,
arnold3} may derive the system toward faster equilibrium.
However, the inclusion of such effects also does not reproduce small 
thermalization time.

The perturbative QCD (pQCD) calculations indicate that the heavy quark ($Q$)
thermalization time, $\tau_i^Q$ is larger~\cite{moore,baier} than the 
light quarks 
and gluons thermalization scale $\tau_i$. 
Gluons may thermalized even before 
up and down quarks~\cite{japrl,shuryak}. In the present work we 
assume that the QGP
is formed at time $\tau_i$. Therefore, the interaction
of the non-equilibrated heavy quarks with the equilibrated QGP for
the time interval $\tau_i<\tau<\tau_i^Q$ can be treated within the ambit 
of the FP equation~\cite{landau,balescu}
{\it i.e.} the heavy quark can be thought of
executing Brownian motion~\cite{moore,japrl,sc,svetitsky,rapp,turbide,
bjoraker,npa1997,munshi,rma} 
in the heat bath of QGP during the said interval of time. 
Therefore, the propagation of a heavy quarks through 
QGP may be treated as the 
interactions between equilibrium and non-equilibrium degrees
of freedom. The FP equation provide an appropriate framework for such
studies. Boltzmann transport equation has recently been applied
to study the depletion of high energy gluons due to its elastic
and inelastic interactions with QGP~\cite{greiner}.

The paper is organized as follows. In the next section the evolution
of the momentum distribution of heavy quarks in QGP are discussed.
In section III we address the issues 
of radiative energy loss with dead cone effect. The non-photonic electron 
spectra is discussed in section IV. The initial conditions and space time 
evolution have been discussed in section V, section VI contains the discussion
on the nuclear  suppression and  finally section VII is 
devoted to summary and conclusions. 

\section{Evolution of heavy quark momentum distributions} 

The Boltzmann transport equation describing a non-equilibrium 
statistical system reads:
\be
\left[\frac{\partial}{\partial t} 
+ \frac{\bf p}{E}.\bf{\nabla_x} 
+ {\bf F}.\bf{\nabla_p}\right]f(x,p,t)=
\left[\frac{\partial f}{\partial t}\right]_{col}
\ee
where $p$ and $E$ denote momentum and energy, ${\bf{\nabla_x}}$
(${\bf{\nabla_p}}$) are spatial (momentum space) gradient and $f(x,p,t)$
is the phase space distribution (in the present case $f$ stands for
heavy quark distribution).
The assumption of uniformity in the plasma and absence of any external force
leads to
\be 
\frac{\partial f}{\partial t}=
\left[\frac{\partial f}{\partial t}\right]_{\mathrm col}
\ee
The collision term on the right hand side of the above equation can be 
approximated as (see ~\cite{svetitsky,npa1997} for details): 
\be
\left[\frac{\partial f}{\partial t}\right]_{col} = 
\frac{\partial}{\partial p_i} \left[ A_i(p)f + 
\frac{\partial}{\partial p_i} \lbrack B_{ij}(p) f \rbrack\right] 
\label{expeq}
\ee
where we have defined the kernels 
\begin{eqnarray}
&& A_i = \int d^3 k \omega (p,k) k_i \nonumber\\
&&B_{ij} = \int d^3 k \omega (p,k) k_ik_j.
\end{eqnarray}
for $\mid\bf{p}\mid\rightarrow 0$,  $A_i\rightarrow \gamma p_i$ 
and $B_{ij}\rightarrow D\delta_{ij}$ where $\gamma$ and $D$ stand for
drag and diffusion co-efficients respectively.
The function $\omega(p,k)$ is given by
\be
\omega(p,k)=g\int\frac{d^3q}{(2\pi)^3}f^\prime(q)v\sigma_{p,q\rightarrow p-k,q+k}
\ee
where $f^\prime$ is the phase space distribution, in the present
case it stands for light quarks and gluons,
$v$ is the relative velocity between the two collision partners,
$\sigma$ denotes the cross section and $g$ is the statistical
degeneracy. The co-efficients in the first two terms of the expansion
in Eq.~\ref{expeq} are comparable in magnitude because the averaging
of $k_i$ involves greater cancellation than the averaging of the
quadratic term $k_ik_j$. The higher power of $k_i$'s are smaller~\cite{landau}.

With these approximations the Boltzmann equation reduces to a non-linear
integro-differential equation known as Landau 
kinetic equation:
\be
\frac{\partial f}{\partial t} = 
\frac{\partial}{\partial p_i} \left[ A_i(p)f + 
\frac{\partial}{\partial p_i} \lbrack B_{ij}(p) f\rbrack \right] 
\label{landaueq}
\ee
The nonlinearity is caused due to the
appearance of $f^\prime$ in $A_i$ and $B_{ij}$ through $w(p,k)$.
It arises from the simple fact that we are studying
a collision process which involves two particles - it should,
therefore, depend on the states of the two participating particles in
the collision process and hence on the product of the two distribution 
functions.
Considerable simplicity may be achieved by replacing the distribution
functions of one of the collision partners by their 
equilibrium Fermi-Dirac or Bose-Einstein distributions
(depending on the statistical nature)
in the expressions of $A_i$ and $B_{ij}$. Then Eq.~\ref{landaueq} 
reduces to a linear
partial differential equation - usually referred to as the FP
equation describing the interaction of a particle which
is out of thermal equilibrium with the particles in a thermal bath
of light quarks, anti-quarks and gluons.
The quantities $A_i$ and $B_{ij}$ are related to the usual 
drag and diffusion coefficients and we denote them by $\gamma_i$ and
$D_{ij}$ respectively ({\it i.e.} these quantities can be obtained
from the expressions for $A_i$ and $B_{ij}$ by replacing the distribution
functions by their thermal counterparts).

The evolution of the heavy quark momentum distribution ($f$) while 
propagating through QGP can 
be studied  by using  the FP equation (see ~\cite{svetitsky} for details)
\be
\frac{\partial f}{\partial t} =
\frac{\partial}{\partial p_i} \left[ \gamma_i(p)f +
\frac{\partial}{\partial p_i} \lbrack D_{ij}(p) f \rbrack \right]
\label{FPeq}
\ee
During the propagation through the QGP the heavy quarks dissipate
energy predominantly by two processes: (i) collisional, {\it e.g.}  
$gQ \rightarrow gQ$,  $qQ \rightarrow qQ$ and $\bar{q}Q \rightarrow \bar{q}Q$
and (ii) radiative processes,  {\it i.e.} when the heavy quark emits 
gluons due to its interaction with the thermal partons in the plasma.
Therefore, the drag and diffusion coefficient should include these two processes
of energy dissipation. 

The elastic collisions of heavy quarks 
with light quarks ($q$)
and gluons ($g$) {\it i.e.}: $gQ \rightarrow gQ$, $qQ \rightarrow qQ$ 
and $\bar{q}Q \rightarrow \bar{q}Q$ 
have been used to evaluate the transport coefficients ($\gamma_{\mathrm coll}$
and $D_{\mathrm coll}$) due to collisional
process.  
At LHC energy one can not ignore the radiative energy loss,
therefore, this should also be taken into account through the transport 
coefficients.  
The transport coefficient~\cite{baier2}, $\hat{q}$,  which is related
to the energy loss~\cite{baier1}, 
$dE/dx$ of the propagating partons in the medium,
has been used to calculate the shear viscosity to entropy 
density ratio, $\eta / s$~\cite{majumder,roy}. 
The  $\hat{q}$ is closely related 
to the diffusion coefficient $D$ (for detail see~\cite{majumder}). 
In similar spirit we use $dE/dx$ to calculate the drag  
coefficient of the medium and use Einsteins relation,
$D=TM\gamma$
to obtain the diffusion co-efficient  when
a heavy quark of mass $M$ is propagating through the
medium at temperature $T$.
The action of drag on the heavy quark can be defined 
through the relation:
\be
-\frac{dE}{dx}\mid_{\mathrm rad}=\gamma_{\mathrm rad}\,p
\ee
where $\gamma_{\mathrm rad}$ denotes the drag-coefficient 
and $p$ is the momentum 
of the heavy quark. It should be mentioned here that the
collisional and the radiative processes are not
entirely independent, {\it i.e.} the collisional 
process may influence the radiative one, therefore strictly speaking 
$dE/dx$ and hence the transport coefficients for
radiative and collisional  process should not be added to obtain the 
net energy loss or net value of the drag coefficient. 
However, in the absence any rigorous way, we add them up
to obtain the effective drag co-efficients, $\gamma_{\mathrm eff}
=\gamma_{\mathrm rad}+\gamma_{coll}$ and similarly the effective
diffusion coefficient: $D_{\mathrm eff}=D_{\mathrm coll}+D_{\mathrm rad}$.
This is a good approximation for the present 
work because the radiative loss is large compared to the collisional loss 
at LHC.
With these effective
transport coefficients the FP equation reads: 
\be
\frac{\partial f}{\partial t} =
\frac{\partial}{\partial p_i} \left[ \gamma_{\mathrm eff}(p)f +
\frac{\partial}{\partial p_i} \lbrack D_{\mathrm eff}(p) f \rbrack \right]
\label{FPeq}
\ee
where $\gamma_{\mathrm eff}$ and $D_{\mathrm eff}$ contain contributions 
from both the mechanisms (collisional and radiative). 
In evaluating the drag co-efficient we have 
used temperature dependent  strong coupling,
$\alpha_s$ from ~\cite{Kaczmarek}.
The Debye mass, $\sim g(T)T$ is  also a temperature dependent
quantity used as  cut-off to shield the infrared divergences
arising due to the exchange of massless gluons.

\section{Energy dissipation processes}
The matrix element for the radiative process ({\it e.g.} $Q+q\rightarrow
Q+q+g$) can be factorized into an elastic process 
($Q+q\rightarrow Q+q$) and a gluon emission ($Q\rightarrow Q+g$).
The emitted gluon distribution can be written as~\cite{gunion,chang}:
\be
\frac{dn_g}{d\eta d^2k_\perp×}=\frac{C_A\alpha_s}{\pi^2} 
\frac{q_\perp^2}{k_\perp^2(k_\perp-q_\perp)^2} F^2
\label{gunion}
\ee
where $k=(k_0,k_\perp,k_3)$ is the four momenta of 
the emitted gluon and $q=(q_0,q_\perp,q_3)$ is the four momenta of 
the exchanged gluon,
$\eta=1/2\,ln(k_0+k_3)/(k_0-k_3)$ is the rapidity and
$C_A=3$ is the Casimir invariant of the adjoint representation and
$\alpha_s=g^2/4\pi$ is the strong coupling constant.

The effects of quark mass in the gluon radiation is taken 
into account by multiplying the emitted gluon 
distribution from massless quarks by $F^2$,
containing the effects of heavy quark mass. 
$F$ is given by~\cite{rkellis,DK}:
\be
F=\frac{k_\perp^2}{\omega^2\theta_0^2+k_\perp^2}
\label{deadcone}
\ee
where $\theta_0=M/E$.

As the energy loss of heavy quark is equal to the energy which is
taken away by the radiated gluon, we can estimate the energy loss of heavy 
quark by multiplying the interaction rate $\Lambda$ and the average energy loss
per collision $\epsilon$, which is given by the average of the probability of 
radiating a gluon times the energy of the gluon. 

The LPM
effects has been taken into account by including a formation time restriction 
on the phase space of the emitted gluon in which the formation time, $\tau_F$
must be smaller than the interaction time, $\tau=\Lambda^{-1}$. The radiative
energy loss of heavy quark can be given by
\be
-\frac{dE}{dx}\mid_{\mathrm rad}=\Lambda\epsilon=\tau^{-1}.\epsilon
\label{dedx}
\ee
where $\epsilon$, the average energy per collision is~\cite{chang,MPST}
\be
\epsilon = \textless n_gk_0 \textgreater=\int d\eta d^2k_\perp 
\frac{dn_g}{d\eta d^2k_\perp×} k_0 \Theta(\tau-\tau_F)F^2
\ee
where $\tau_F={\mathrm cosh}\eta/k_\perp$. 
As mentioned before for the infrared cut-off $k^{min}_\perp$ we choose the Debye 
screening mass of  gluon. 
\be
k_\perp^{min}=\mu_D=\sqrt{4\pi \alpha_s}T
\ee
The maximum transverse momentum of the emitted 
gluon is given by:
\begin{eqnarray}
(k_\perp^{max})^2=\textless \frac{(s-m^2)^2}{4s×}\textgreater= 
\frac{3ET}{2×}-\frac{m^2}{4×} \nonumber \\+\frac{m^4}{48pT×}
ln[\frac{m^2+6ET+6pT}{m^2+6ET-6pT×}]
\end{eqnarray}

\begin{figure}[h]
\begin{center}
\includegraphics[scale=0.43]{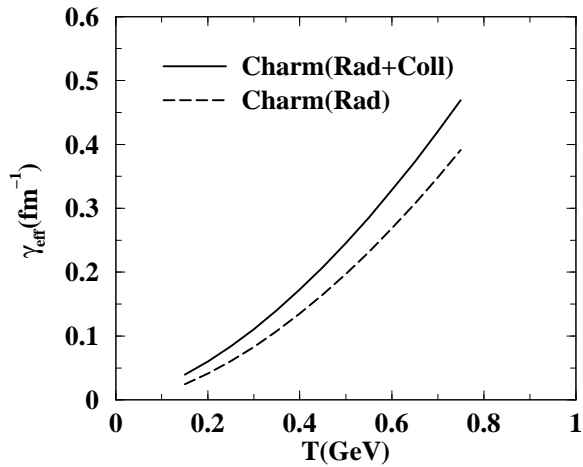}
\caption{Variation of effective drag coefficient with temperature 
for charm quarks}
\label{fig1}
\end{center}
\end{figure}

\begin{figure}[h]
\begin{center}
\includegraphics[scale=0.43]{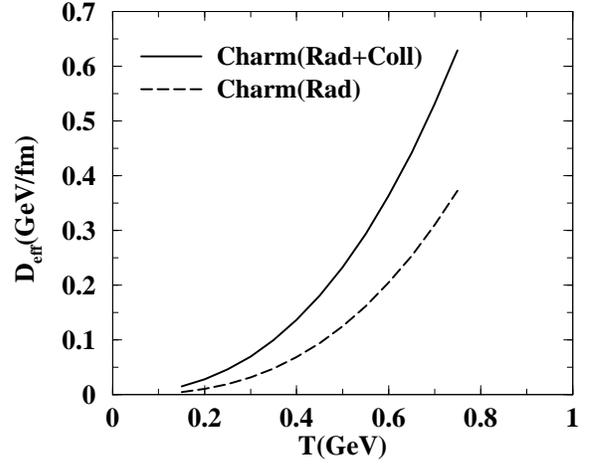}
\caption{Variation of effective diffusion coefficient with temperature
for charm quarks }
\label{fig2}
\end{center}
\end{figure}

\begin{figure}[h]
\begin{center}
\includegraphics[scale=0.43]{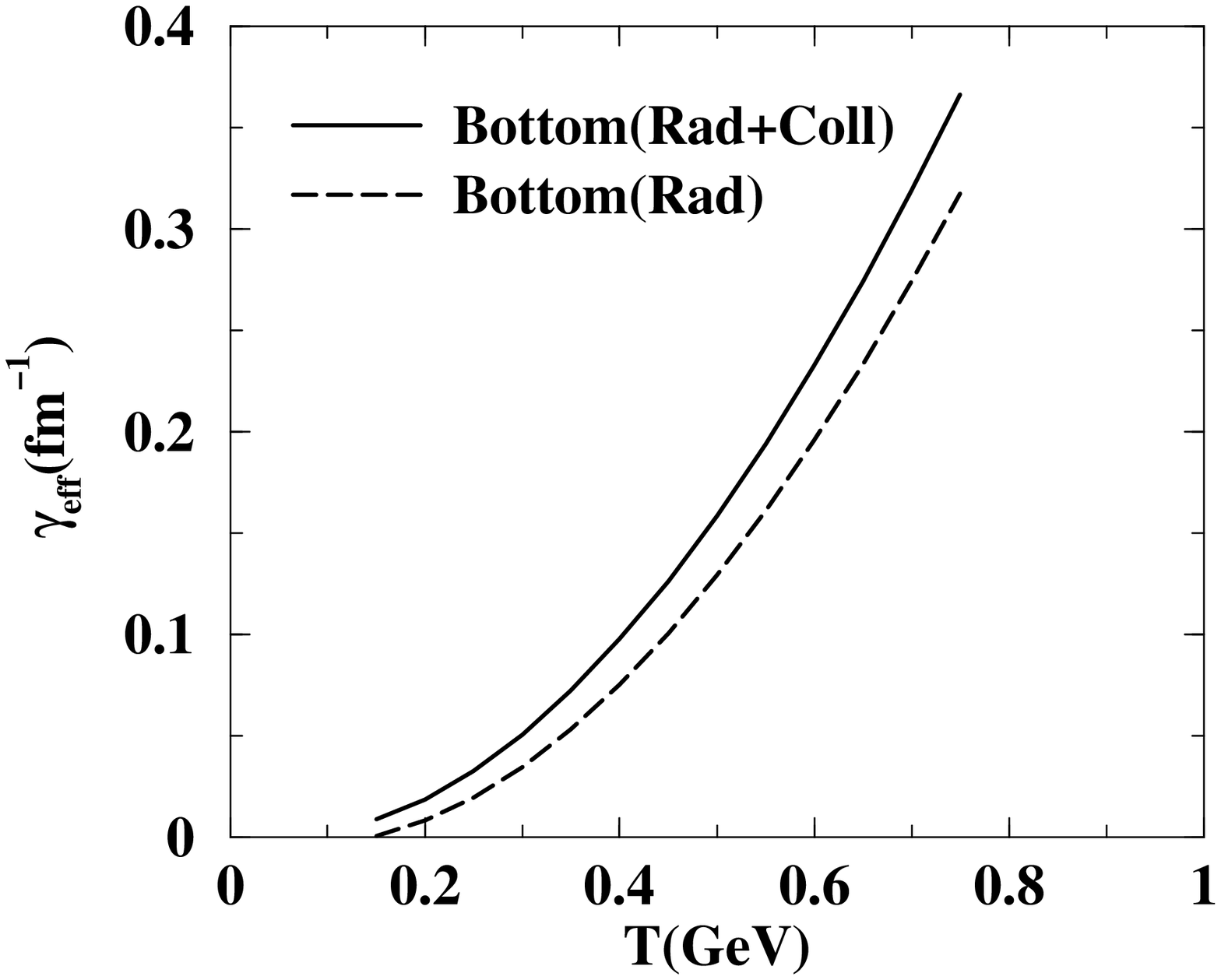}
\caption{Same as Fig.~{\protect\ref{fig3}} for bottom quarks}
\label{fig3}
\end{center}
\end{figure}

\begin{figure}[h]
\begin{center}
\includegraphics[scale=0.43]{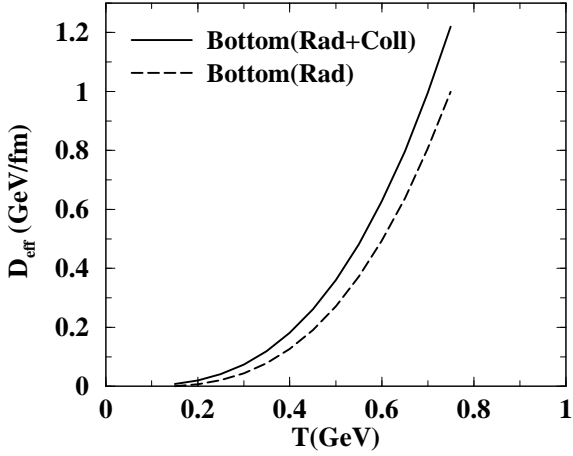}
\caption{Same as Fig.~{\protect\ref{fig2}} for bottom quarks}
\label{fig4}
\end{center}
\end{figure}
Following the procedure of earlier works~\cite{svetitsky,skdas}
we evaluate the drag and diffusion coefficients for the 
elastic processes.  Knowing $\gamma_{\mathrm rad}$ from the
radiative processes as described above we obtain the effective
drag coefficients and hence effective diffusion coefficient
through Einstein relation.
In Figs.~\ref{fig1} and ~\ref{fig2} 
the variation of effective drag and diffusion coefficients 
with $T$ have been depicted for charm quarks. We observe that 
the contribution of the radiative loss is large compared to the collisional or
elastic one. The difference between the collisional and radiative
loss increases with temperature - indicating very small contribution
from the former at large $T$. 
Similar difference is reflected in the diffusion coefficients
as we have used Einstein's relation to obtain it from the drag coefficients.
We observe that at low $T$ and $p_T$ the contributions from collisional
processes is more than or comparable to that from radiative processes. 
For the bottom quark we find that the gap between the drag coefficients
with radiative and elastic processes 
is  smaller (compared to charm) at lower temperature domain.
Quantitatively the value of drag is smaller
for bottom than charm quarks because of their larger relaxation time.
However, the 
qualitative behaviour is similar to charm quarks as shown in Fig.~\ref{fig3}.
The diffusion coefficient of the bottom quark is large (~Fig.\ref{fig4})
compared to the charm quark because of the large mass of the former 
introduced  through the Einstein's relation.  

On obtaining 
the effective drag and diffusion coefficients next we need to know the initial 
heavy quarks momentum distributions to solve the FP equation.
The production of charm and
bottom quarks in hadronic collisions is studied extensively~\cite{charmbottom}.
In the present work the $p_T$ distribution of 
charm and bottom quarks in pp collisions have been taken from the 
NLO MNR code~\cite{MNR}. The results  from the code
may be tested by measuring the production cross sections of
heavy mesons (containing $c$ and $b$  quarks) in pp collisions at 
$\sqrt{s_{\mathrm NN}}=5.5$ TeV. With all these required inputs
we solve the FP equation by using the Greens function technique 
(see~\cite{skdas,rapp} for details).

\section{THE NON-PHOTONIC ELECTRON SPECTRA}
The FP equation has been solved for the heavy quarks
with the initial condition mentioned above. 
We convolute the solution with the fragmentation functions of the
heavy quarks to obtain the $p_T$ distribution of the
heavy mesons ($B$ and $D$) ($dN^{D,B}/q_Tdq_T$). For heavy quark 
fragmentation 
we use Peterson function~\cite{peterson} given by:
\be
f(z) \propto 
\frac{1}{\lbrack z \lbrack 1- \frac{1}{z}- \frac{\epsilon_c}{1-z} \rbrack^2 \rbrack}
\ee
for charm quark $\epsilon_c=0.05$. For bottom quark 
$\epsilon_b=(M_c/M_b)^2\epsilon_c$ where $M_c$ ($M_b$) is the charm
(bottom) quark mass.
The non-photonic single electron spectra originate from the
decays of heavy flavoured mesons - {\it e.g.} $D\rightarrow Xe\nu$
or $B\rightarrow Xe\nu$
at mid-rapidity ($y=0$) can be obtained as follows~\cite{gronau,ali}:
\be
\frac{dN^e}{p_Tdp_T}=\int dq_T \frac{dN^D}{q_Tdq_T} F(p_T,q_T)
\ee
where
\be
F(p_T,q_T)=\omega\int \frac{d(\bf{p}_T.\bf{q}_T)}{2p_T\bf{p}_T.\bf{q}_T}g(\bf{p}_T.\bf{q}_T/M)
\ee
where $M$ is the mass of the heavy mesons ($D$ or $B$),
$\omega=96(1-8m^2+8m^6-m^8-12m^4lnm^2)^{-1}M^{-6}$ ($m=M_X/M$) and $g(E_e)$ is given by
\be
g(E_e)=\frac{E_e^2(M^2-M_X^2-2ME_e)^2}{(M-2E_e)}
\ee
related to the  rest frame spectrum for the decay $D\rightarrow X e \nu$
through the following relation~\cite{gronau}
\be
\frac{1}{\Gamma_H}\frac{d\Gamma_H}{dE_e}=\omega g(E_e).
\ee

We evaluate the electron spectra from the decays of heavy mesons
originating from the fragmentation of the heavy quarks propagating
through the QGP formed in heavy ion collisions.
Similarly the electron spectrum from the p-p collisions
can be obtained from the charm and bottom quark distribution
which goes as the initial conditions to the solution of FP equation.
The ratio of these two quantities, $R_{AA}$ then gives,
\be
R_{AA}(p_T)=\frac{\frac{dN^e}{d^2p_Tdy}^{\mathrm Au+Au}}
{N_{\mathrm coll}\times\frac{dN^e}{d^2p_Tdy}^{\mathrm p+p}}
\label{raa}
\ee
called the nuclear suppression factor,
will be unity in the absence of any medium. In Eq.~\ref{raa}
$N_{\mathrm coll}$ stands for the number of nucleon-nucleon 
interactions in a nucleus+nucleus collision. The experimental
data ~\cite{stare,phenixelat,phenixe} at RHIC energy ($\sqrt{s_{NN}}$=200 GeV)
shows substantial suppression ($R_{AA}<1$) for $p_T\geq 2$ GeV indicating
substantial interaction of the plasma particles with charm and bottom quarks
from which electrons are originated through the process:
$c(b)$ (hadronization)${\longrightarrow}$ $D(B)$(decay)$\longrightarrow$
$e+X$. The loss of energy of high momentum heavy quarks propagating through
the medium created in Au+Au collisions causes a depletion of high $p_T$
electrons.

\section{Space time evolution}
The system formed in nuclear collisions at relativistic energies
evolves dynamically from the initial to the final state. 
The time evolution such systems may be studied 
by solving the hydrodynamic equations:
\be
\partial_\mu T^{\mu\nu}=0
\ee
with boost invariance along the longitudinal
direction~\cite{jdbjorken}. 
In the above equation $T^{\mu\nu}=(\epsilon+P)u^\mu u^\nu-g^{\mu\nu}P$, 
is the energy momentum tensor for ideal fluid, 
$\epsilon$ is the energy density, $P$ is the pressure
and $u^\mu$ is the hydrodynamic
four velocity. It is expected that the central rapidity
region of the system formed after nuclear collisions
at LHC energy is almost net baryon free. Therefore,
the equation governing the conservation of net
baryon number need not be considered here. 
The radial co-ordinate dependence of $T$ 
have been parametrized as in Ref.~\cite{turbide}.
Some comments on the effects
of the radial flow are in order here. The radial
expansion will increase the size of the system and
hence decrease the density of the medium. Therefore,
with radial flow the heavy quark will traverse
a larger path length in a medium of reduced density.
These two oppositely competing phenomena may
have negligible net effects on the nuclear 
suppression(see also ~\cite{turbide}).

The total amount of energy dissipated by a  heavy quark in the QGP
depends on the path length it traverses.
Each parton traverse different path length
which depends on the  geometry of the system and on the point 
where its is created.
The probability that a parton is produced at a point $(r,\phi)$
in the plasma depends on the number of binary collisions 
at that point which can be taken as:
\be
P(r,\phi)=\frac{2}{\pi R^2}(1-\frac{r^2}{R^2})\theta(R-r)
\label{prphi}
\ee
where $R$ is the nuclear radius. It should be mentioned here
that the expression in Eq.~(\ref{prphi}) is an approximation for the
collisions with zero impact parameter.
A parton created at $(r,\phi)$ in the transverse plane
propagate a distance $L=\sqrt{R^2-r^2sin^2\phi}-rcos\phi$
in the medium. In the present work we use the following
equation for the geometric average of the integral
involving drag coefficient [$\int d\tau\gamma(\tau$)]: 
\be
\Gamma=\frac{\int rdr d\phi P(r,\phi) \int^{L/v}d\tau\gamma(\tau)}
{\int rdr d\phi P(r,\phi)}
\label{cgama}
\ee
where $v$ is the velocity of the propagating partons. 
Similar averaging has been performed  for the diffusion co-efficient.
For a static system the temperature dependence of the drag and
diffusion co-efficients of the heavy quarks enter via the
thermal distributions of light quarks and gluons through
which it is propagating. However, in the present scenario
the variation of temperature with time is governed by
the equation of state or velocity of sound
of the thermalized system undergoing hydrodynamic
expansion. In such a scenario the quantities like $\Gamma$ (Eq.~\ref{cgama})
and hence $R_{\mathrm AA}$ becomes sensitive to velocity of sound ($c_s$)
in the medium. This will be shown in the next section.

\section{THE NUCLEAR SUPPRESSION}
The $p_T$ dependence of   $R_{AA}$ is sensitive to the 
nature of the initial (prior to the interaction with the 
medium) distribution of heavy quarks~\cite{MNR}. 
For the QGP expected to be formed at the
LHC we have taken initial temperature, $T_i=700$ MeV, initial
thermalization time $\tau_i=0.08$ fm/c which reproduces the
predicted hadron multiplicity $dN/dy=2100$~\cite{armesto} through the relation:
\be
T_i^{3}\tau_i \approx \frac{2\pi^4}{45\zeta(3)}\frac{1}{4a_{eff}}\frac{1}
{\pi R_A^2}\frac{dN}{dy}.
\label{eq6}
\ee
where
$R_A$ is the radius of the system,
$\zeta(3)$ is the Riemann zeta function and  $a_{eff}=\pi^2g_{eff}/90$ where
$g_{eff}$ ($=2\times 8+ 7\times 2\times 2\times 3\times N_F/8$) is the
degeneracy of quarks and gluons in QGP, $N_F$=number of flavours
We have taken the value of the transition temperature, $T_c=170$ MeV.

The value of $R_{AA}$ is plotted against the $p_T$ of
the non-photonic single electron resulting from $D$ decays
in Fig~\ref{fig5}. The results show substantial depletion at
large $p_T$ indicating large interaction rate of the charm 
quarks with the thermal medium of partons. The sensitivity of the 
results on the equation of state is also demonstrated in Fig.~\ref{fig5}.
A softer equation of state (lower value of $c_s$) makes the expansion of
the plasma slower, enabling the propagating heavy quarks to
spend more time interacting in the medium and hence lose
more energy before exiting from the plasma, which results
in less particle production at high $p_T$. This is clearly demonstrated in
Fig.~\ref{fig5}. It may be mentioned here that $c_s$ increases with 
temperature. Therefore, due the higher initial temperature of
the QGP formed at LHC the value of $c_s$ 
may be larger than that of QGP formed at RHIC energies. Keeping this
in mind  we predict the nuclear suppression factors for three values of 
$c_s=1/\sqrt{3}$ (maximum possible), 
$1/\sqrt{4}$ and $1/\sqrt{5}$ (~Fig.\ref{fig5}). 

The nuclear suppression for the bottom quarks are displayed in 
Fig.~\ref{fig6}. We observe quantitatively less suppression 
compare to  charm quarks. The difference between the 
charm and bottom quarks suppression 
are affected chiefly by two factors:
(i) for different values transport coefficients and
(ii) for the different kind of initial $p_T$ distributions. 
The bottom  quark has less drag coefficients and 
has  harder $p_T$ 
distributions  - both these factors are responsible for the 
smaller suppression of bottom quark.    
The present results on $R_{\mathrm AA}$  may be compared with those 
obtained in~\cite{Gyulassy2005} in a different approach.

\begin{figure}[h]
\begin{center}
\includegraphics[scale=0.43]{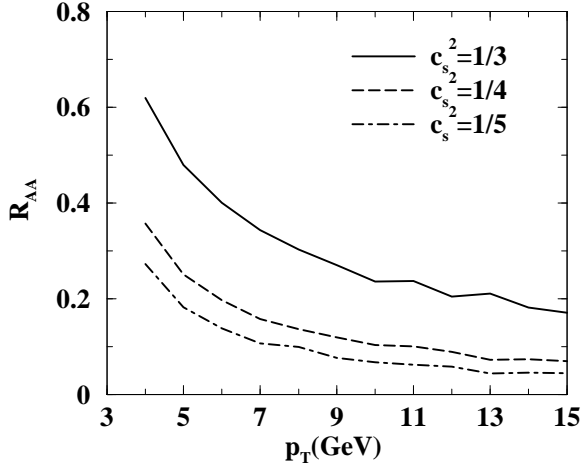}
\caption{Nuclear suppression factor, $R_{\mathrm AA}$ as a function of $p_T$
for various equation of state for non-photonic single
electron resulting form $D$-mesons decay.}
\label{fig5}
\end{center}
\end{figure}

\begin{figure}[h]
\begin{center}
\includegraphics[scale=0.43]{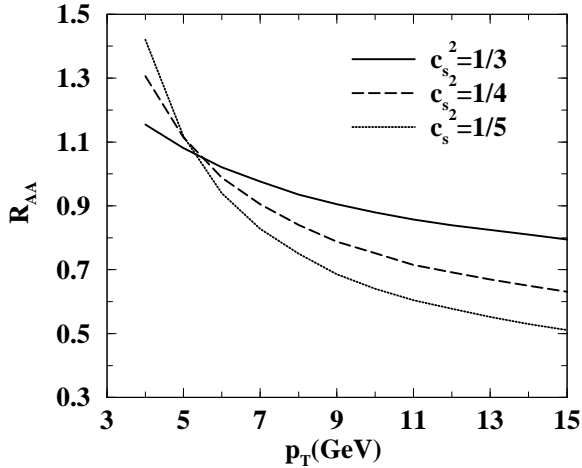}
\caption{Same as Fig.~{\protect\ref{fig5}} for $B$ mesons.
} 
\label{fig6}
\end{center}
\end{figure}

\begin{figure}[h]
\begin{center}
\includegraphics[scale=0.43]{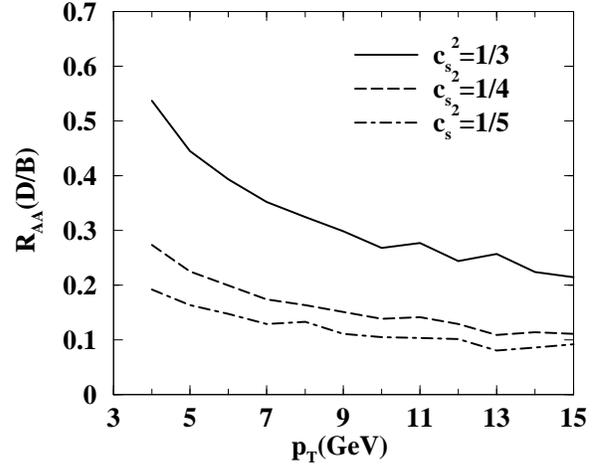}
\caption{
Variation of the ratio of nuclear suppression factor, $R_{\mathrm AA}$ 
for charm to bottom quarks as a function of $p_T$. 
}
\label{fig7}
\end{center}
\end{figure}
\begin{figure}[h]
\begin{center}
\includegraphics[scale=0.43]{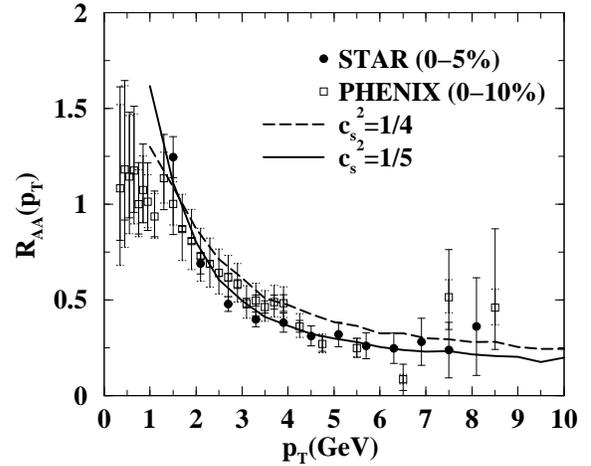}
\caption{Comparison of $R_{\mathrm AA}$ obtained in the present work
with the experimental data obtained by STAR and PHENIX collaboration
for $\sqrt{s_{\mathrm NN}}=200$ GeV. The experimental data of STAR
and PHENIX collaborations are 
taken from ~\cite{stare} and~\cite{phenixelat} respectively.
 }
\label{fig8}
\end{center}
\end{figure}
In Fig~\ref{fig7} we have plotted the ratio: 
$R_{\mathrm AA}^D/R_{\mathrm AA}^B$ as a function of $p_T$,
from where the effect of the mass and the role of
the nature (soft or hard) of the initial $p_T$ distributions
can be understood (see also~\cite{cacciari}).

In Fig.~\ref{fig8} we compare the experimental data obtained by  the
STAR~\cite{stare} and PHENIX~\cite{phenixelat} collaborations 
for Au + Au collisions at $\sqrt{s_{\mathrm NN}}=200$ GeV with
theoretical results obtain in the present work. For the 
theoretical calculations the value of initial and transition temperatures are
taken as 400 MeV and 170 MeV respectively. The value of
initial thermalization time is assumed as 0.2 fm/c.
These values of initial thermalization time and initial
temperature reproduces the total multiplicity at mid-rapidity,
$dN/dy= 1100$.
We observe that the data can reasonably be reproduced by 
taking velocity of sound $c_s=1/\sqrt{5}$. 
It should be mentioned here that the inclusion of both radiative and
elastic losses in the effective drag enables us to reduce the 
gap between the experiment and theory without any enhancement of the pQCD 
cross section as has been done in our previous work~\cite{skdas}.

So far we have discussed the suppression of the
non-photonic electron produced in nuclear collisions
due to the propagation of the heavy quark in the 
partonic medium in the pre-hadronization era. However,
the suppression of the D mesons in the post
hadronization era (when both the temperature and
density are lower than the partonic phase) should in
principle be also taken into account. The suppression of the D 
mesons in the post hadronization era is found  to
be small~\cite{skdas1}, indicating the fact that
the hadronic medium (of pions and nucleons) is unable
to drag the $D$ mesons strongly.

\section{Summary and Conclusions}
We have evaluated the drag and diffusion co-efficients containing
both the elastic and radiative loss for charm and bottom quarks.
We found that the radiative loss is dominant over its collisional 
counterpart. In the radiative process dead cone and LPM effects are
taken in to account.
With these  transport coefficients and initial
charm and bottom $p_T$ distributions from NLO MNR~\cite{MNR} code 
we have solved the FP equation. The solution of FP
equation has been used to predict nuclear suppression factors to be 
measured through the semi-leptonic decays of heavy 
mesons ($D$ and $B$) for LHC conditions.  We find that 
the suppression is quite large indicating that the 
heavy quarks undergo substantial interactions in the 
QGP medium.  The ratio of the suppression for $D$ and $B$ quarks
has also been evaluated to understand the effects of mass on the suppression. 
The same formalism has been applied to study the experimental data
on non-photonic single electron spectra measured by STAR and PHENIX 
collaborations at the highest RHIC energy. The data is well reproduced 
without any enhancement of the pQCD cross section.

{\bf Acknowledgment:}
We are grateful to Matteo Cacciari for providing us the heavy 
quarks transverse momentum distribution and also for useful discussions.
This work is supported by DAE-BRNS project Sanction No. 2005/21/5-BRNS/2455.

\end{document}